\documentclass[preprint,12pt]{elsarticle}

\usepackage{graphicx}
\usepackage{amssymb}

\begin{document}

\begin{frontmatter}

\title{Anomalous Transport Processes in Turbulent non-Abelian Plasmas}

\author{Masayuki Asakawa}
\address{Department of Physics, Osaka University, Toyonaka 560-0043, Japan}

\author{Steffen A.~Bass}
\address{Department of Physics, Duke University, Durham, NC 27708, USA}

\author{Berndt M\"uller}
\address{Department of Physics \& CTMS, Duke University, Durham, NC 27708, USA}

\begin{abstract}
Turbulent color fields, which can arise in the early and late stages of relativistic heavy ion collisions, may contribute significantly to the transport processes in the matter created in these collisions. We review the theory of these {\em anomalous} transport processes and discuss their possible phenomenology in the glasma and quasistationary expanding quark-gluon plasma.
\end{abstract}

\end{frontmatter}

\section{Introduction}
\label{sec:intro}

The theory of perturbative saturation of strong glue fields (the ``color glass condensate'') in fast moving nuclei provides compelling reasons to believe that the pre-equilibrium stage of nuclear collisions at high energy is governed by the nonlinear dynamics of color fields at moderate coupling $\alpha_s(Q_s^2)$, where $Q_s \sim 1-2$ GeV is the nuclear saturation scale \cite{McLerran:2008es}. Recent improved simulations of lattice QCD are compatible with a quasiparticle structure of QCD matter at thermal equilibrium (the quark-gluon plasma) even at temperatures as low as 250 MeV \cite{Cheng:2009zi}. This suggests that the matter created in heavy ion collisions at top RHIC energy is -- at least initially -- not as strongly coupled as many aspects of the phenomenology of these reactions, {\em viz.}~the near maximal elliptic flow and the strong jet quenching, seem to indicate \cite{Gyulassy:2004zy}.

It is thus worthwhile asking the question whether the extreme opaqueness of the quark-gluon plasma observed in the RHIC experiments can be explained without invoking a super-strong coupling? Here we argue that the answer may lie in the peculiar transport properties of turbulent non-Abelian plasmas.

\section{Plasma Turbulence}
\label{sec:turb}

The term {\em plasma turbulence} describes a random, nonthermal pattern of excitation of coherent field modes in a plasma with a power spectrum similar to that of vortices in a turbulent fluid \cite{turbulence}. This phenomenon is usually caused by plasma instabilities. One class of such instabilities, which exists in both Abelian and non-Abelian plasmas, is the instability discovered by Weibel \cite{Weibel:1959}. It is driven by the interaction of soft gauge field modes with the particles that constitute the plasma. In the Weibel instability certain gauge field modes spontaneously start to grow at an exponential rate whenever the momentum distribution of charged particles is anisotropic. An extreme case of this instability is the famous two-stream instability, which was first studied for non-Abelian plasmas by Mr\'owczy\'nski \cite{Mrowczynski:1996vh}.

Such Weibel-type instabilities arise naturally in expanding quark-gluon plasmas, because the width of the momentum component in the direction of the expansion narrows by dilution, leading to an anisotropic momentum distribution. The extreme cases are the free-streaming Bjorken scenario, in which the width of the longitudinal momentum distribution shrinks as $1/\tau$ while the transverse momentum distribution remains fixed, and the scenario of viscous hydrodynamics, where the plasma stays near equilibrium, but the longitudinal momentum spread is less than the transverse momentum spread by an amount proportional to the shear viscosity $\eta$ \cite{Martinez:2010sc}:
\begin{equation}
\frac{1}{2}\langle {\bf p}_T^2\rangle - \langle p_L^2\rangle\approx \frac{4\pi^2}{3}\frac{\eta}{s}\frac{T}{\tau} .
\end{equation}
If it persists long enough, the instability develops into full-scale plasma turbulence through mode coupling  resulting in a power-law excitation spectrum of soft gauge field modes \cite{Arnold:2005vb,Rebhan:2005re}.

The other instability type, first discovered by Nielsen and Olesen \cite{Nielsen:1978rm} only occurs in non-Abelian plasmas. It is caused by the strong interaction between the gluon spin and chromo-magnetic fields, which is attractive when the spin aligns with the field. Fujii and Itakura \cite{Fujii:2008dd} and independently Iwazaki \cite{Iwazaki:2008xi} recently showed that the Nielsen-Olesen (NO) instability also occurs in longitudinally expanding chromo-magnetic flux tubes, which are predicted to be formed as a result of the interaction of two color glass condensate states in a relativistic heavy ion collision. The dynamic evolution of these glasma field configurations was studied numerically in \cite{Romatschke:2005pm,Rebhan:2008uj}.

\section{Anomalous Transport}
\label{sec:anom}

As we have shown \cite{Asakawa:2006tc,Asakawa:2006jn}, soft color fields generate anomalous transport coefficients which may dominate the transport properties of the plasma at weak and moderately weak coupling. The two most relevant transport coefficients are the shear viscosity $\eta$ and the jet quenching parameter $\hat{q}$. The latter is proportional to the mean squared momentum per unit length imparted by the turbulent fields on an energetic parton; the former is inversely proportional to the same quantity (for partons of ``average'' momentum). 

\begin{figure}
\label{fig:anomvisc}
\centerline{\includegraphics[width=0.6\textwidth]{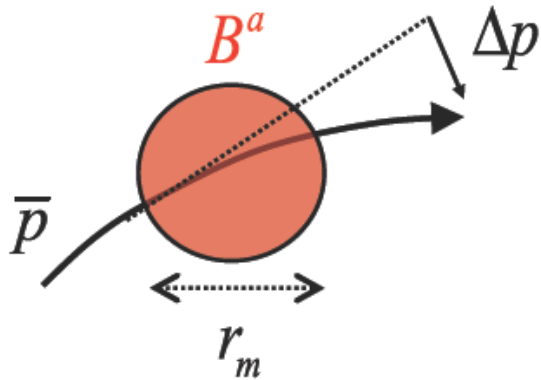}}
\caption{Schematic diagram showing the deflection $\Delta{\bf p}$ of a perturbative parton of momentum $\bar{p}$ traversing a coherent domain of chromomagnetic field ${\bf B}^a$.}
\end{figure}

To understand this, consider Fig.~\ref{fig:anomvisc}. The classical expression for the shear viscosity from kinetic theory is
\begin{equation}
\eta = \frac{1}{3} n\, \bar{p}\, \lambda_f ,
\end{equation}
where $n$ is the density of the medium, $\bar{p}$ is the mean momentum of particles and $\lambda_f$ is the distance over which their momentum becomes randomized by interactions in the medium (the ``mean free path''). If the medium is filled with domains of locally coherent gauge fields, the momentum of particles change due to deflection by these fields. Let us assume that the fields $B^a$ are of chromo-magnetic nature and have a coherence length $r_m$. We also denote the color charge of a particle by $Q_a$. Then the momentum change incurred by a particle crossing one of these domains is
\begin{equation}
\label{eq:Deltap}
\Delta p = g\, Q^aB^a \, r_m .
\end{equation}
The distance after which the particle momentum has been randomized is then given by
\begin{equation}
\lambda_f = r_m\,  \left\langle \bar{p}^2/(\Delta p)^2 \right\rangle 
= \frac{\bar{p}^2}{g^2\, Q^2 \langle B^2 \rangle\, r_m} .
\end{equation}
The {\em anomalous} shear viscosity of the plasma is thus given by
\begin{equation}
\eta_{\rm A} = \frac{n\, \bar{p}^3}{3 g^2\, Q^2 \langle B^2 \rangle\, r_m} 
\approx \frac{\frac{9}{4} s\, T^3}{g^2\, Q^2 \langle B^2 \rangle\, r_m} ,
\end{equation}
where the last expression is valid if the plasma is near equilibrium. Here $s\approx 4n$ denotes the entropy density of the quark-gluon plasma.

A similar calculation applies to the jet quenching parameter $\hat{q}$, which is a measure of the transverse momentum diffusion of a fast parton as it propagates through the turbulent quark-gluon plasma:
\begin{equation}
\hat{q} = \frac{\left\langle (\Delta p_\perp(L))^2 \right\rangle}{L} .
\end{equation}
Again, the momentum change in one coherent domain is given by (\ref{eq:Deltap}), resulting in the {\em anomalous} contribution to the jet quenching parameter:
\begin{equation}
\hat{q}_{\rm A} = \frac{\left\langle (\Delta p_\perp)^2 \right\rangle}{r_m}
= g^2\, Q^2 \langle B^2 \rangle\, r_m .
\end{equation}
In the quasithermal medium, this implies the relationship \cite{Majumder:2007zh}
\begin{equation}
\label{eq:eta-qhat}
\frac{\eta_{\rm A}}{s} \approx c\, \frac{T^3}{\hat{q}_{\rm A}} ,
\end{equation}
where $c$ is a constant of order unity. The relation (\ref{eq:eta-qhat}) is generally valid in gauge field plasmas, where transport processes are dominated by small angle scattering.

\section{Anomalous Transport Theory}
\label{sec:anomtheory}

The formal theory of anomalous transport processes \cite{Asakawa:2006jn} starts from the Vlasov-Boltzmann equation for the phase space distribution of particles:
\begin{equation}
\label{eq:VB}
\left[ \frac{\partial}{\partial t} + \frac{\bf p}{E_p}\cdot\nabla_r + {\bf F}\cdot\nabla_p \right] 
f({\bf r},{\bf p}; t) = C[f] ,
\end{equation}
where $C[f]$ denotes the collision term and $\bf F$ is the local color force
\begin{equation}
{\bf F} = g\, Q^a \left( {\bf E}^a + {\bf v}\times{\bf B}^a \right) .
\end{equation}
Assuming that the chromo-electric and -magnetic fields ${\bf E}^a$ and ${\bf B}^a$ are random, (\ref{eq:VB}) can be transformed into a Fokker-Planck equation, which describes the diffusion of plasma particles in momentum space:
\begin{equation}
\label{eq:ABM}
\left[ \frac{\partial}{\partial t} + \frac{\bf p}{E_p}\cdot\nabla_r - \nabla_p\cdot D({\bf v})\cdot\nabla_p \right] 
f({\bf r},{\bf p}; t) = C[f] ,
\end{equation}
where the collision term can be neglected if one is only interested in the anomalous transport processes. The diffusion coefficient is given by the time integral over the force correlation function along the trajectory of a plasma particle:
\begin{equation}
D_{ij}({\bf v};{\bf r},t) = \int_{-\infty}^t dt'\, \left\langle F_i({\bf r}(t'),t') F_i({\bf r},t) \right\rangle .
\end{equation}
Here ${\bf r}(t')$ is the position at time $t'$ of the particle which arrives at position ${\bf r}$ at time $t$. The diffusion constant $D_{ij}$ is the generalization of the expression $g^2\, Q^2 \langle B^2 \rangle\, r_m$, which occurs in the heuristic expressions for the anomalous shear viscosity and jet quenching parameter.

\section{Discussion}
\label{sec:disc}

For an expanding almost equilibrated quark-gluon plasma, the anomalous shear viscosity can be shown to dominate over the collisional shear viscosity for a fixed velocity gradient in the weak coupling limit  \cite{Asakawa:2006jn}. Applying a self-consistency condition -- that the momentum anisotropy is governed by the anomalous shear viscosity which, in turn, is controlled by the turbulent plasma fields generated by the gauge field instabilities of the anisotropic plasma -- one finds the scaling law
\begin{equation}
\label{eq:etas-A}
\frac{\eta_{\rm A}}{s} \sim \left( \frac{T}{g^3\, |\nabla u|} \right)^\nu ,
\end{equation} 
where $\nu \approx \frac{1}{2}$. Here $|\nabla u|$ is the magnitude of the velocity gradient. On the other hand, the collisional shear viscosity scales as \cite{Arnold:2000dr}
\begin{equation}
\label{eq:etas-C}
\frac{\eta_{\rm C}}{s} \sim \frac{1}{g^4\,\ln g^{-1}} .
\end{equation} 
Since $\eta = ( \eta_{\rm A}^{-1} + \eta_{\rm A}^{-1} )^{-1}$, the anomalous shear viscosity dominates the total shear viscosity for a given velocity gradient at weak coupling $g \to 0$. Where the transition between the anomalous and the collisional regime occurs depends on the numerical constants missing in the scaling relations (\ref{eq:etas-A}, \ref{eq:etas-C}). The momentum diffusion constant has not been systematically evaluated as a function of the momentum anisotropy of the turbulent plasma, but a value can in principle be deduced from numerical simulations of momentum broadening of hard partons \cite{Dumitru:2007rp,Schenke:2010}. 

In the {\em glasma} phase, the situation is different from the hydrodynamic regime, because most of the energy density is in the form of coherent color fields. As a consequence, anomalous transport mechanisms are bound to dominate over collisional (Boltzmann) transport processes. An estimate for the anomalous jet quenching parameter in the glasma can be obtained as 
\begin{equation}
\hat{q}(\tau) \approx \frac{Q_s^3}{Q_s\tau} \approx \frac{10~{\rm GeV}^2/{\rm fm}}{Q_s\tau} ,
\end{equation}
which is in reasonable agreement with experimentally deduced values of $\hat{q}$ extrapolated to early times. More accurate determinations of the momentum diffusion constant in the glasma phase by numerical simulations of the nonlinear gauge field dynamics \cite{Krasnitz:2001qu,Lappi:2003bi} would be of acute interest.

{\em Acknowledgment:} This work was supported in part by the U.S. Department of Energy under grant DE-FG02-05ER41367.

\bibliographystyle{elsarticle-num}

\end{document}